\documentclass[aps,twocolumn]{revtex4}
\usepackage{epsf}
\usepackage{bm}

\begin{document}

\title{Possibility for exciton Bose-Einstein condensation in carbon nanotubes}

\author{I. V. Bondarev and A. V. Meliksetyan}

\affiliation{Department of Math \& Physics, North Carolina Central University, Durham, NC 27707, USA}

\begin{abstract}
We demonstrate a possibility for exciton Bose-Einstein condensation in individual small-diameter ($\sim\!1\!-\!2$~nm) semiconducting carbon nanotubes.~The effect occurs under the exciton-interband-plasmon coupling controlled by an external electrostatic field applied perpendicular to the nanotube axis. It requires fields $\sim\!1$~V/nm and temperatures below $100$~K that are experimentally accessible. The effect offers a testing ground for fundamentals of condensed matter physics in one dimension and opens up perspectives to develop tunable coherent polarized light source with carbon nanotubes.
\end{abstract}

\maketitle

\section{Introduction}

Carbon nanotubes (CNs), graphene sheets rolled-up into cylinders of one to a few nanometers in diameter and up to hundreds of microns in length, have been successfully integrated into miniaturized electronic, electromechanical, chemical devices, scanning probes, and into nanocomposite materials~\cite{Dresselhaus01,Baughman13}. Over the past few years, optical nanomaterials research has uncovered intriguing optical attributes of their physical properties~\cite{ChemPhysSI}, as well, lending themselves to a variety of optoelectronic device applications~\cite{ChemPhysSI,Avouris08,McEuen09,Hertel10,Mueller10,Malic11,Bondarev12,Bondarev12pss,Kamat07,Strano11,Kono12,Dresselhaus07}. The great breadth and depth of optical phenomena in CNs is exemplified by experimental and theoretical reports on how their optical properties are affected by defects~\cite{Hartschuh05,Rotkin12}, exciton-phonon interactions~\cite{Ferrari07}, biexciton formation~\cite{Pedersen05,Bondarev11}, exciton-plasmon coupling~\cite{Bondarev09}, external magnetic~\cite{Kono08} and electric fields~\cite{Bondarev09,Perebeinos07}. Recent studies have also looked at thermal rectification~\cite{Zettle06}, microwave-frequency signal rectification~\cite{Cobas08} and non-linear optical response of individual CNs~\cite{Nasibulin11}. Several recent efforts reported on semiconducting CNs used for the generation, detection and harvesting of light~\cite{McEuen09,Hertel10,Mueller10,
Malic11,Bondarev12,Bondarev12pss,Kamat07,Strano11}, and as single photon sources for quantum computing, communication, or cryptography~\cite{Xia08,Hoegele,Bondarev10}.

Undoped semiconducting single-wall CNs are direct band-gap semiconductors and feature very large exciton binding energies (hundreds of meV)~\cite{Dresselhaus07,Avouris08}, which one can vary in controllable ways to modify their underlying optical properties. This offers new functionality and creates a very strong potential for future tunable optoelectronic device applications with individual CNs. Excitons in CNs can be affected by either electrostatic doping~\cite{Mueller10,Spataru10} or by the quantum confined Stark effect (QCSE)~\cite{Bondarev09,Bondarev12} (an external electrostatic field applied perpendicular to the CN axis; the effect is used recently to tune the bandgap of bilayer graphene~\cite{Fengwang09}). In both cases, exciton properties are mediated by collective plasmon excitations.~QCSE, in particular, allows one to control exciton-interband-plasmon coupling in individual undoped CNs and their (linear~\cite{Bondarev09,Bondarev12,Bondarev12pss} and nonlinear~\cite{Bondarev11,Bondarev11pss}) optical absorption~\cite{endnote}.~By varying the exciton-plasmon coupling strength with the QCSE one controls both the radiative exciton emission and non-radiative exciton-to-plasmon energy transfer. This latter phenomenon is similar to the SPASER effect (Surface Plasmon Amplification by Stimulated Emission of Radiation) reported earlier for hybrid metal-semiconductor-dielectric nanostructures~\cite{Stockman}.~It takes place in \emph{individual} small-diameter CNs though, resulting in new strongly coupled hybridized excitations --- exciton-plasmons --- and associated high-intensity coherent oscillating fields concentrated locally along the CN surface~\cite{Bondarev12,Bondarev12pss}. These near-fields can be used in a variety of new optoelectronic applications, including near-field nonlinear-optical probing and sensing, optical switching, enhanced electromagnetic absorption, and materials nanoscale modification.

Apart from applications, carbon nanotubes offer an ideal testing ground to study the fundamentals of condensed matter physics in one dimension (1D). Here, we discuss possibilities for the 1D Bose-Einstein condensation (BEC) phenomenon that originates from the strong coupling of excitons and inter-band (same-band) plasmons enabled by using the QCSE. Exciton-plasmons thus created in an individual nanotube are strongly correlated collective Bose excitations and, therefore, could likely be condensed under appropriate external conditions --- in spite of the well-known statements of the BEC impossibility in ideal 1D/2D systems~\cite{Feynman} and experimental evidence for no exciton BEC effect in highly excited semiconducting CNs~\cite{Kono09}. Possibilities for achieving BEC in 1D and 2D systems are theoretically demonstrated in the presence of an extra confinement potential~\cite{Kleppner91}. We show that the strongly correlated exciton-plasmon system in a CN in an external perpendicular electrostatic field presents such a special case.~We find the critical BEC temperature, as well as the condensate fraction and its exciton contribution as functions of temperature and electrostatic field applied.~We discuss how the effect can be observed experimentally.

\section{Dispersion of Exciton-Plasmons}

\begin{figure}[t]
\epsfxsize=9.0cm\centering{\epsfbox{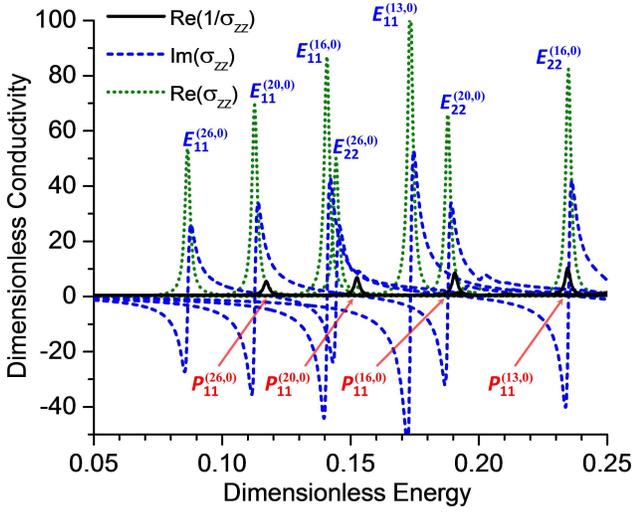}}
\caption{%
(Color online) Calculated energy dependence of the dimensionless (normalized by $e^2/2\pi\hbar$) axial surface conductivity $\sigma_{zz}$ for the four zigzag nanotubes, (13,0), (16,0), (20,0) and (26,0), of increasing diameters.~Peaks of $\mbox{Re}\,\sigma_{zz}$ represent excitons ($E_{11}$, $E_{22}$); peaks of $\mbox{Re}(1/\sigma_{zz})$ represent inter-band plasmons ($P_{11}$, $P_{22}$).~Dimensionless energy is defined as $[Energy]/2\gamma_0$, where $\gamma_0=2.7$~eV is the C-C overlap integral.}
\label{fig1}
\end{figure}

\begin{figure}[b]
\epsfxsize=8.5cm\centering{\epsfbox{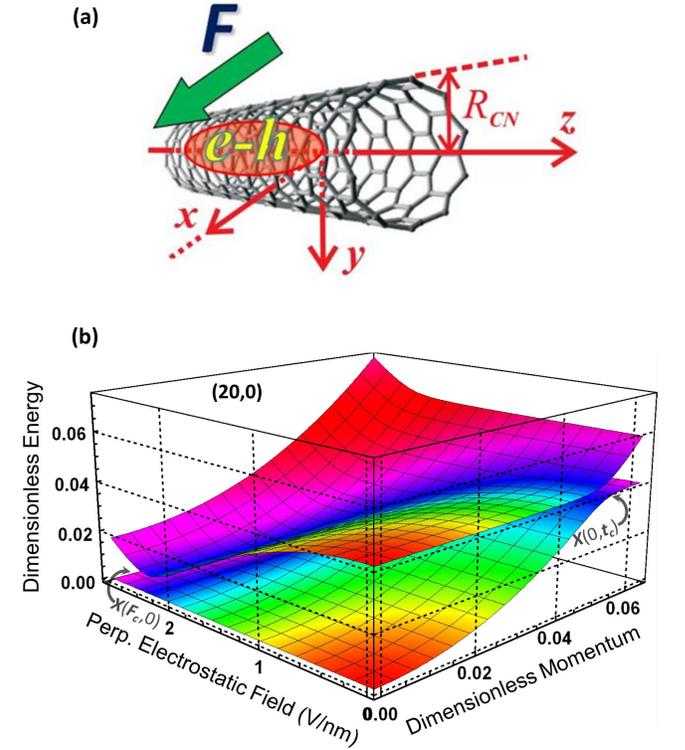}}\vspace{-0.4cm}
\caption{%
(Color online) (a)~The geometry of the problem. (b)~Exciton-plasmon dispersion as a function of the perpendicular electrostatic field and longitudinal momentum for the lowest bright exciton coupled to the nearest inter-band plasmon in the (20,0) nanotube ($E_{11}^{(20,0)}$ and $P_{11}^{(20,0)}$ in Fig.~\ref{fig1}). See text for dimensionless momentum.}
\label{fig2}\vspace{-0.4cm}
\end{figure}

Figure~\ref{fig1} shows our calculations of the dynamical axial conductivities for a few representative semiconducting CNs, (13,0), (16,0), (20,0) and (26,0), of increasing diameters $\sim\!1\!-\!2$~nm ($R_{CN}\!=\!0.51, 0.63, 0.78~\mbox{and}~1.02$~nm, respec\-tively).~In small-diameter CNs, excitons are excited by the external electromagnetic (EM) radiation polarized along the CN axis~\cite{Ando}.~So, only the axial conductivity $\sigma_{zz}$ matters while the azimuthal one, $\sigma_{\varphi\varphi}$, can be neglected being strongly suppressed due to the transverse depolarization effect [we use cylindrical coordinates, Fig.~\ref{fig2}~(a), with the $z$-axis being the CN axis]. To calculate $\sigma_{zz}$, we used the $(\textbf{k}\cdot\textbf{p})$-method of Ref.~\cite{Ando} with the exciton relaxation time 100~fs for all four CNs (consistent with previous estimates~\cite{Ferrari07,Perebeinos07}).~Many-particle Coulomb correlations are included by solving the Bethe-Salpeter equation in the momentum space within the screened Hartree-Fock approximation as described in Ref.~\cite{Ando}. Real conductivities consist of series of peaks ($E_{11},E_{22},...$) representing the 1st, 2nd, etc., excitons. Imaginary conductivities are linked with the real ones by the Kramers-Kronig relation.~That is why the functions $\mbox{Re}(1/\sigma_{zz})\!=\!\mbox{Re}(\sigma_{zz})/\{[\mbox{Re}(\sigma_{zz})]^2+[\mbox{Im}(\sigma_{zz})]^2\}$ show the resonances $P_{11},P_{22},...$ right next to $E_{11},E_{22},...~$. These are the inter-band plasmon resonances (first studied experimentally in Ref.~\cite{Pichler98}) as $\mbox{Re}(1/\sigma_{zz})$ is directly related to $\mbox{Im}(1/\epsilon_{zz})$, the electron-energy-loss spectroscopy response ($\epsilon_{zz}$ is the longitudinal dielectric function) representative of plasmon excitations in the system~\cite{Pichler98}.

Perpendicular electrostatic field [Fig.~\ref{fig2}~(a)] mixes exciton and plasmon resonances, to result in two branches of new hybridized quasi-particle states, exciton-plasmons, with the energies (dimensionless variables, see Ref.\cite{Bondarev09})
\begin{equation}
x_{1,2}=\sqrt{\frac{\varepsilon_f^2+x_{p}^2}{2}\pm\frac{1}{2}
\sqrt{(\varepsilon_f^2\!-x_{p}^2)^2+(2X_f)^2\varepsilon_fx_{\!p}}}\;,
\label{x12}
\end{equation}
where $\varepsilon_f=E_f(\mathbf{k})/2\gamma_0$ and $x_p=E_p/2\gamma_0$ are the energies of the $f$-internal-state exciton and inter-band plasmon of the same band, respectively, with $\gamma_0=2.7$~eV being the C-C overlap integral, $X_f=[2\Delta x_{p}\bar\Gamma_0^f(x_{p})\rho(x_{p})]^{1/2}$ is the exciton-plasmon Rabi-splitting taken at energy $x=x_p$, where $\bar\Gamma_0^f$ is the exciton spontaneous decay (radiative recombination) rate and $\rho(x)\!\approx\!\rho(x_p)\Delta x_{p}^2/[(x-x_{p})^2+\Delta x_{p}^2]$ is the Lorentzian approximated (of half-width-at-half-maximum $\Delta x_p$ which is proportional to the inverse plasmon life-time) density-of-states (DOS) representing the non-radiative coupling of excitons to plasmons. Exciton quasi-momentum ${\bf k}\!=\!\{k_\varphi,k_z\}$ is a two-component vector with quantized $k_\varphi$ and continuous $k_z$ to represent the longitudinal motion of the exciton center-of-mass, so that $E_f(\mathbf{k})=E_{exc}^{(f)}(k_{\varphi})+\hbar^2k_z^2/2M_{ex}(k_{\varphi})$ with the first term $E_{exc}^{(f)}(k_{\varphi})=E_g(k_{\varphi})+E_b^{(f)}(k_{\varphi})$ being the exciton excitation energy \{$E_g$ is the band gap, $E_b^{(f)}$ is the (negative) binding energy; these, as well as $x_p$, are affected by the QCSE, see Ref.\cite{Bondarev09}\} and the second term being the kinetic energy of the translational longitudinal motion of the exciton with the effective mass $M_{ex}=m_e+m_h$, where $m_{e,h}$ are the electron (hole) effective masses.

Figure~\ref{fig2}~(b) shows an example of the two exciton-plasmon branches given by Eq.~(\ref{x12}) as functions of the perpendicular electrostatic field $F$ [geometry shown in Fig.~\ref{fig2}~(a)] and longitudinal momentum for the lowest bright ground-internal-state exciton (in which case we drop the $f$-subscript in what follows) coupled to the nearest inter-band plasmon in the (20,0) nanotube [$E_{11}^{(20,0)}$ and $P_{11}^{(20,0)}$ in Fig.~\ref{fig1}].~The origin of the energy is taken to be $x_2(F,k_z=0)$.~In non-zero field $F_c\!\approx\!2$~V/nm, where the strong exciton-plasmon coupling occurs, the upper branch $x_1$ has the global minimum at zero momentum, $k_z^c\!=\!k_z(F_c)\!=\!0$, separated from the lower branch~by~the Rabi-splitting $X(F_c,0)=X(F_c)$.~Hence, at equilibrium, if the temperature $T$ is such that $k_BT/2\gamma_0\!>\!X(F_c)$, strongly coupled upper-branch exciton-plasmons will be distributed around this minimum in the momentum space.~Lowering $T$ to get $k_BT/2\gamma_0\!<\!X(F_c)$ will push them all down to occupy the lowest possible energy state, the $k_z\!=\!0$ state, which is nothing but the exciton-plasmon BEC effect.~This effect does not depend on the density of particles though, as opposed to the BEC of non-interacting massive bosons~\cite{Feynman,Landau}.~Rather, this is the characteristic feature of the quasi-particle energy spectrum, its dependence on the electrostatic field applied, to be exact, of the coupled exciton-plasmon excitations. Therefore, this BEC effect is hardly sensitive to the interaction, if any, between exciton-plasmons in our system.

\begin{figure}[t]
\epsfxsize=8.5cm\centering{\epsfbox{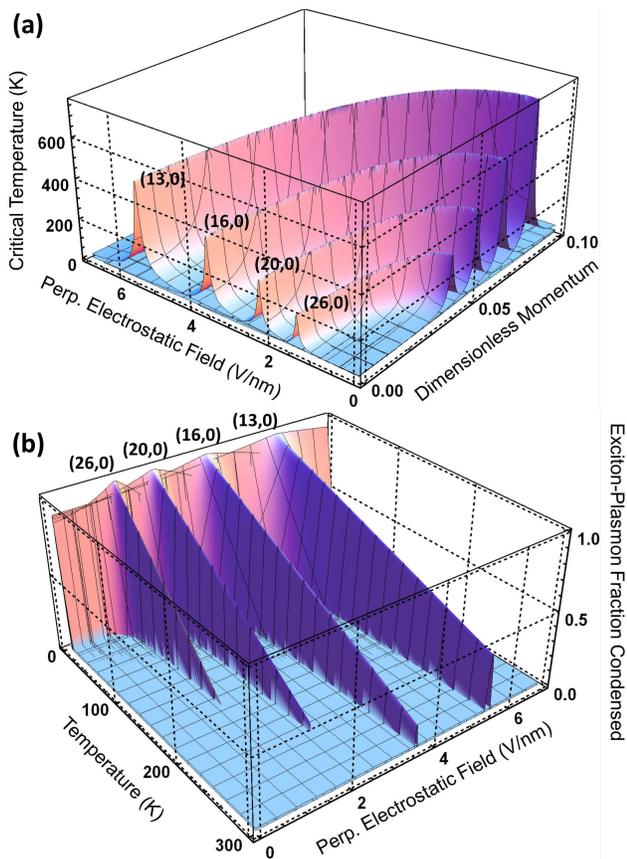}}\vspace{-0.4cm}
\caption{%
(Color online) Critical temperatures (a) as functions of the perpendicular electrostatic field applied and longitudinal momentum, and mean upper-branch BEC population fractions (b) as given by Eq.~(\ref{n10}), for the four CNs under consideration. See text for dimensionless momentum.}
\label{fig3}\vspace{-0.4cm}
\end{figure}

\section{Exciton BEC effect}

It is not difficult to derive the BEC fraction as a function of $T$ and $F$. At $k_z\!\sim\!k_z^c=0$, wherein $\varepsilon_f\!\sim\!x_p$, Eq.~(\ref{x12}) expands into $x_1(F,s,t)\approx X_f(F,s)+\alpha(s)\,t^2/2$,~with the energy counted from $x_2(F,s,t\!=\!0)$, $t\!=\!k_z/\tilde{k}_z$ being the di\-men\-sion\-less longitudinal quasi-momentum ($|t|\!\le\!1$) and $\alpha=\hbar^2\tilde{k}_z^2/2M_{ex}(s)$.~The first Brillouin zone of the CN of $(m,n)$ type ($n\le m$) is taken to be consisting of $m$ parallel lines, as per quantized $k_{\varphi}=k_{\varphi}(s)=s/R_{C\!N}$ with $s=1,2,...,m$ and $R_{C\!N}=(\sqrt{3}\,b/2\pi)\sqrt{m^2+mn+n^2}$, each of length $2\tilde{k}_z=2B/k_{\varphi}(m)$, where $2B=2(4\pi^2/3\sqrt{3}\,b^2)$ is the rectangular area of the reciprocal space covered by the lines, $b=1.42$~\AA~ is the C-C interatomic distance~\cite{Bondarev12}. To obtain the upper-branch exciton-plasmon mean BEC population fraction $\langle n_{10}\rangle$, we use this $x_1$ and employ the conventional technique (e.g.,~Refs.~\cite{Feynman,Landau}) to perform the summation over $\mathbf{k}$ in the first Brillouin zone. For the ground-internal-state exciton, assuming $M_{ex}(s)\!\approx\!M_{ex}$ and $X_f(F,s)\!\approx\!X(F)$, this results in (see Appendix~\ref{appA} for details)
\begin{equation}
\langle n_{10}\rangle(T\!\le\!T_c,F)\approx1-T/T_c(F),
\label{n10}
\end{equation}
where $T_c(F)\!=\!2\gamma_0X(F)/k_B$ is the critical temperature with $X(F)\!=\!\{2\Delta x_{p}\bar\Gamma_0[\varepsilon(F,t\!=\!0)]\rho[\varepsilon(F,t\!=\!0)]\}^{1/2}$ standing for the exciton-plasmon Rabi-splitting at the (excitation) energy of the zero-momentum ground-state exciton. This latter one controls the convergence of the result.

\begin{figure}[b]
\epsfxsize=8.5cm\centering{\epsfbox{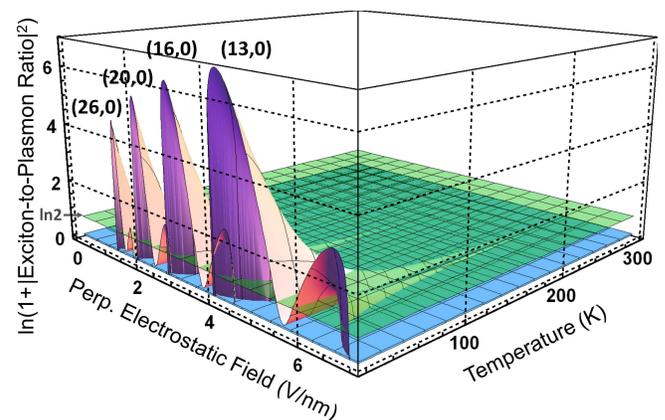}}\vspace{-0.4cm}
\caption{%
(Color online) Function $\ln[1+\langle n_{10}\rangle_{\!_{E\!/\!P}}(0,T,F)]$ calculated using Eqs.~(\ref{n10}) and (\ref{n10ep}) for the CNs under consideration.}
\label{fig4}\vspace{-0.4cm}
\end{figure}

Figure~\ref{fig3}~(a) shows calculated $T_c$ as functions of $F$ and~$t$ (to better understand the general behavior), for the lowest bright ground-internal-state excitons coupled to the nearest inter-band plasmons in the four CNs under consideration.~The functions $T_c(F,t)$~are resonance shaped of widths $\sim\!2\Delta x_p$, peaked at $\varepsilon(F,t)\!\approx\!x_p(F)$.~As $F$ increases, the peak positions shift down to $t\!\sim\!0$, yielding field dependent, resonance shaped $T_c(F)$, same as $X(F)$, peaked at $F_c\sim\!1\!-\!6$~V/nm with maximum $T_c\!\sim\!150\!-\!500$~K and greater $F_c$ and $T_c$ for smaller diameter CNs.~Figure~\ref{fig3}~(b) presents $\langle n_{10}\rangle$ as given by Eq.~(\ref{n10}) for the same case.~The quantities $\langle n_{10}\rangle$ reflect the behavior of $T_c(F)$, showing finite BEC fractions throughout the finite ranges of $F$ centered about $F_c$, expanding in $F$ as $T$ decreases.

Following the general theory of the exciton-plasmon interactions in individual CNs~\cite{Bondarev09}, one can now calculate the exciton participation rate in the exciton-plasmon BEC population fraction (\ref{n10}). This is represented by the absolute value squared of the ratio of the exciton mixing coefficient to the plasmon mixing coefficient of the upper exciton-plasmon branch. Using the mixing coefficients of Ref.\cite{Bondarev09} yields thus defined exciton participation rate as a function of dimensionless energy $x$, $T$ and $F$ as follows (Appendix~\ref{appB})
\begin{equation}
\langle n_{10}\rangle_{\!_{E\!/\!P}}(x,T,F)\approx\frac{\pi\Delta x_p(x-x_1)^2(1+x_1/\varepsilon)^2}{X^2(F)}\langle n_{10}\rangle
\label{n10ep}
\end{equation}
with energies to be counted from $x_2(F,t\!=\!0)$.~Comparing this with Fig.~\ref{fig2}~(b), we see that at $x\!=\!0$ corresponding to the first ground-state-exciton excitation energy and $F\!=\!F_c$ [resonance condition yielding $x_1\!=\!X(F_c)$ and $\varepsilon\!=\!X(F_c)/2$], Eq.~(\ref{n10ep}) becomes $9\pi\Delta x_{\!p}\langle n_{10}\rangle\!\ll\!1$,~meaning that the exciton-plasmon BEC is dominated by plasmons. This is consistent with the coherent plasmon generation effect by excitons reported lately~\cite{Bondarev12,Bondarev12pss}. A slight detuning from $F_c$ increases $x_1$ and dramatically decreases $X(F)$ [Fig.~\ref{fig3}~(a)], bringing about bursts of $\langle n_{10}\rangle_{\!_{E\!/\!P}}\!\gg\!1$, making the exciton-plasmon BEC dominated by excitons.

The exciton BEC effect is shown in Fig.~\ref{fig4} for the CNs under consideration. It occurs at $F\!\gtrsim\!1$~V/nm and $T\!\lesssim\!100$~K that are experimentally accessible~\cite{Fengwang09}. The effect is stronger and covers broader range of $F$ and $T$ in smaller diameter CNs. Off-resonance coupling to inter-band plasmons, which are just \emph{standing} charge density waves, slows excitons down by pushing them into periodic effective potential "traps", thereby increasing zero-momentum exciton state population. This is consistent with earlier studies where the 1D/2D BEC phenomenon, otherwise prohibited~\cite{Feynman,Hohenberg,Pitaevskii}, is shown to occur in the presence of an extra confinement potential~\cite{Kleppner91}.

\section{Conclusions}

The effect of the exciton BEC presented here will manifest itself as highly coherent, longitudinally polarized, far-field exciton emission appearing at temperatures below 100~K as one smoothly increases the perpendicular field strength.~Narrow BEC emission peak will be blue shifted by the Rabi-splitting energy from the first exciton excitation energy which the CN should be pumped at by an external laser source.~The phenomenon can be investigated in experiments similar to those used for exciton-polariton BEC studies in semiconductor microcavities~\cite{Littlewood}.

To summarize, we have demonstrated theoretically an intriguing possibility for the exciton BEC effect observation in individual semiconducting carbon nanotubes. The quantum system considered here is conceptually similar to the microcavity exciton-polariton system, which started as a theoretical concept in the nineties and has been a driving force for the experimental physics of low-dimensional semiconductors over the last two decades, exhibiting both new fundamental quantum effects and attractive applications such as polariton lasers, optical polarization switches, superfluid spintronic devices, etc.~\cite{Kavokin07} We, therefore, strongly believe that the quasi-1D exciton BEC effect predicted here not only offers an ideal testing ground for fundamentals of condensed matter physics in one dimension, but also opens up new horizons for a variety of CN based applications ranging from controlled electromagnetic absorption and tunable highly coherent polarized light emission, in particular, to the extension of nanoplasmonics and near-field optics research, currently focused on metallic nanoparticles~\cite{StockmanPT,NovotnyPT}, to include a new area of nanotube plasmonics. Further new avenues for experimental nanotube plasmonics research could potentially emerge from the measurements of the coherent BEC exciton emission predicted here if experiments will show that the origin of the emission spot can be translated at will by applying, say, a temperature gradient along the nanotube.~Then, a new fundamental effect, the superfluidity of quasi-1D exciton-plasmons in individual CNs, would become a hot research topic for the future. We strongly believe this is only the beginning of nanotube plasmonics as a new research field. Physical understanding of 1D nanophotonics phenomena will greatly benefit from studies of the model system considered herein.

\acknowledgments

Discussions with J.~Anders of UCL, UK, are acknowledged. I.V.B. is supported by DOE (DE-SC0007117); A.V.M. is funded by ARO (W911NF-11-1-0189).

\appendix

\section{~Derivation of Eq.~(\ref{n10})}\label{appA}
We assume our system being under constant illumination by low-intensity external monochromatic radiation polarized along the CN axis. This excites excitons and not plasmons since plasmons are longitudinal excitations while photons are transverse~\cite{Cardona,Toyozawa}.~Under a perpendicular electrostatic field applied, excitons couple to the nearest (same-band) interband plasmons to form new hybridized excitations --- exciton-plasmons~\cite{Bondarev09,Bondarev12,Bondarev12pss}, whose occupation numbers, by virtue of the linear response theory in the exciton-radiation interaction and the corresponding fluctuation dissipation theorem~\cite{Landau}, can be found as an equilibrium statistical average of the form
\begin{equation}
\langle n_\mu(\mathbf{k})\rangle=\mbox{Tr}\!\left[\frac{e^{-\beta\hat{H}}}{Q}\;\hat{\xi}^\dag_\mu(\mathbf{k})\hat{\xi}_\mu(\mathbf{k})\right]\!.
\label{n}
\end{equation}
Here, $\hat{\xi}^\dag_\mu(\mathbf{k})$ and $\hat{\xi}_\mu(\mathbf{k})$ create and annihilate, respectively, the exciton-plasmon excitation of branch $\mu\,(=\!1,2)$ with the momentum $\mathbf{k}$, $\hat{H}\!=\!\sum_{\mathbf{k},\mu=1,2}\hbar\omega_\mu(\mathbf{k})\hat{\xi}^\dag_\mu(\mathbf{k})\hat{\xi}_\mu(\mathbf{k})$ is~the total Hamiltonian diagonalized of the interacting excitons and plasmons, and $\hbar\omega_\mu(\mathbf{k})$ is the exciton-plasmon energy given by the solution to the dispersion relation resulted from the diagonalization. This is given by Eq.~(\ref{x12}) in dimensionless variables $x_\mu=\hbar\omega_\mu/2\gamma_0$ (see Ref.\cite{Bondarev09}). The partition function $Q=\mbox{Tr}(e^{-\beta\hat{H}})$ with $\beta=1/k_BT$. The chemical potential of the exciton-plasmons is zero as consistent with the fact of no mass added to our system by either exciton or plasmon excitation.

The partition function in Eq.~(\ref{n}) can be easily evaluated using the exciton-plasmon occupation number Hilbert space vector set $\prod_{\mu=1}^2\prod_\mathbf{k}|n_\mu(\mathbf{k})\rangle$, resulting in
\begin{equation}
Q=\prod_{\mu=1}^2\prod_\mathbf{k}\frac{1}{1-e^{-\beta\hbar\omega_\mu(\mathbf{k})}}\,.
\label{Q}
\end{equation}
Equation (\ref{n}) can be evaluated in the same manner, using this partition function, to bring us to the massless boson type exciton-plasmon occupation number
\begin{equation}
\langle n_\mu(\mathbf{k})\rangle=\frac{1}{e^{\beta\hbar\omega_\mu(\mathbf{k})}-1}=\frac{1}{e^{\lambda x_\mu(\mathbf{k})}-1}
\label{nk}
\end{equation}
with $\lambda=2\gamma_0\beta$.

To correctly evaluate the mean population fraction for zero-momentum ($k_z\!=\!0$) exciton-plasmons of the upper branch ($\mu\!=\!1$), we follow the conventional procedure of textbook statistical physics (see, e.g., Refs.~\cite{Feynman,Landau}). We begin with the calculation of the upper branch exciton-plasmon mean population in the first Brillouin zone. The first Brillouin zone of the carbon nanotube of $(m,n)$ type ($n\!\le\!m$) is taken to be consisting of $m$ parallel lines, as per quantized $k_{\varphi}=k_{\varphi}(s)=s/R_{C\!N}$ with $s=1,2,...,m$~and $R_{C\!N}=(\sqrt{3}\,b/2\pi)\sqrt{m^2+mn+n^2}$ ($b=1.42$~\AA~being the C-C distance), each of length $2\tilde{k}_z=2B/k_{\varphi}(m)$, where $2B=2(4\pi^2/3\sqrt{3}\,b^2)=(2\pi)^2/2S_0$ is the rectangular area of the reciprocal space covered by the lines and $S_0$ is the equilateral triangle area selected around each C atom in such a way as to cover the entire CN surface~\cite{Bondarev09,Bondarev12}. This yields $\tilde{k}_z=2\pi/3b$ for the $(m,0)$ type CNs (zigzag) and $\tilde{k}_z=2\pi/\sqrt{3}\,b$ for the $(m,m)$ type CNs (armchair), in particular. Since the total number of states in the first Brillouin zone is $N/2$ (two C atoms per elementary cell of a graphene layer of $N$ atoms to form a single wall CN of length $L$, that is $N\!=\!2\pi R_{CN}L/S_0$), the mean population $n_1$ for the upper branch exciton-plasmons is
\[
n_1=\frac{2}{N}\sum_{s=1}^m\sum_{k_z=-\tilde{k}_z}^{\tilde{k}_z}\!\!\!\langle n_1(s,k_z)\rangle
\]\vspace{-0.5cm}
\[
=\frac{2}{N}\sum_{s=1}^m\frac{L}{\pi}\!\int_0^{\tilde{k}_z}\!\!\!\!\!dk_z\,\langle n_1(s,k_z)\rangle\,,
\]
where the fact that $\langle n_1(s,-k_z)\rangle\!=\!\langle n_1(s,k_z)\rangle$ is taken~into account. This can now be rewritten in dimensionless variables to take the form
\begin{equation}
n_1=\frac{1}{m}\sum_{s=1}^m\int_0^{1}\!\frac{dt}{e^{\lambda x_1(s,t)}-1}
\label{n1T}
\end{equation}
with $t\!=\!k_z/\tilde{k}_z$ and $x_1(s,t)$,~the upper branch exciton-plasmon energy, to be taken in~the form of the expansion of Eq.~(\ref{x12}) near $t_c(F)\!=\!0$ under the strong exciton-plasmon coupling condition (controlled by the external perpendicular electrostatic field $F$ by means of the QCSE). Under this condition, $\varepsilon_f\!\sim\!x_p$ [avoided crossing in Fig.~\ref{fig2}~(b)] and the coupling term (that $\sim\!X_f$) is dominant under the square root in Eq.~(\ref{x12}), to result in the expansion as follows
\begin{equation}
x_1(F,s,t)\approx X_f(F,s)+\frac{\alpha(s)}{2}\,t^2
\label{x1fst}
\end{equation}
with $\alpha(s)\!=\!\hbar^2\tilde{k}_z^2/[2M_{ex}(s)2\gamma_0]$ and the energy counted from $x_2(F,s,t\!=\!0)$.

The integral over $t$ in Eq.~(\ref{n1T}) can be done by the (geometric) series expansion with subsequent term-by-term integration. Under the assumption of $M_{ex}(s)\!\approx\!M_{ex}$ and $X_f(F,s)\!\approx\!X(F)$, where $M_{ex}$ and $X(F)$ are those for the first bright exciton in its ground internal state, this results in
\begin{equation}
n_1=\frac{\sqrt{\pi}}{2}\sum_{n=0}^{+\infty}\frac{e^{-(n+1)\lambda X}\,\mbox{Erf}\,[\sqrt{(n+1)\lambda\alpha/2}\,]}{\sqrt{(n+1)\lambda\alpha/2}}\,.
\label{n1TF}
\end{equation}
Although this model assumption might seem to be questionable from the theoretical viewpoint, the experiment can be set up in such a way as to maintain the system continuously illuminated at the first exciton excitation energy. That would correspond to no summation over $s$ and no $1/m$ statistical factor present in the initial Eq.~(\ref{n1T}), since only the first exciton would be there contributing to the entire effect with the statistical factor of one. This brings us back to Eq.~(\ref{n1TF}) again.

The series in Eq.~(\ref{Dn1TF}) is seen to converge due to the presence of the exponential factor decaying at a rate that depends on $F$ and $T$ as the summation index $n$ increases. By d'Alembert ratio test the series is convergent absolutely and uniformly when $\exp(-\lambda X)\!<\!1$. This is always the case at finite $T$ ($=\!2\gamma_0/k_B\lambda$) for non-zero $X$ [$=\!X(F)$], and the greater is $X$, the faster is the convergence. The less is $X$, on the other hand, the slower is the convergence, yielding eventually $\exp(-\lambda X)\!\sim\!1$ at $X\!\sim\!0$ corresponding to divergent harmonic type series typical of 1D and 2D geometries where no BEC phenomenon is known to occur in ideal boson gas type systems~\cite{Feynman,Hohenberg,Pitaevskii}. The quantity $\exp(\lambda X)$ represents the radius of convergence (to be greater than one). This allows us to single out a range of parameters $F$ and $T$ at which the convergence occurs.~Setting up $T_c(F)\!=\!2\gamma_0X(F)/k_B$ yields $F$ such that $\exp(\lambda X)\!=\!\exp[T_c(F)/T]\!>\!1$ for all $T\!\le\!T_c(F)$, so that the series is manifestly convergent.

For $T\!\le\!T_c(F)$ the sum in Eq.~(\ref{n1TF}) can be evaluated using an approximate expression
\begin{equation}
\sum_{n=0}^{\infty}F(n+a)\approx\!\!\int_0^{\infty}\!\!\!\!\!\!dx\,F(x)-F(a)\!\left(\!a-\frac{1}{2}\!\right)+\frac{F^{\prime}(a)}{2}\!\left(\!a^2\!-\frac{1}{6}\!\right)
\label{summationformula}
\end{equation}
that comes from the Euler-Maclaurin summation formula (see Ref.~\cite{Landau})
\[
\frac{F(a)}{2}+\sum_{n=1}^{\infty}F(n+a)\approx\!\int_a^{\infty}\!\!\!\!\!dx\,F(x)-\frac{F^{\prime}(a)}{12}
\]
after rearranging the first two terms of the series on the left, writing $\int_a^\infty\!\!=\!\int_0^\infty\!\!-\!\int_0^a$ on the right and then using a fully legitimate approximation $F(x)\!\approx\!F(a)+F^{\prime}(a)(x-a)$ in the second integral. With $F(n+a)\!=\!F(n+1)$, as per Eq.~(\ref{n1TF}), this brings us to the following result
\begin{equation}
n_1\approx\frac{\arctan\!\left(\!\sqrt{\alpha/2X}\right)}{\lambda\sqrt{\alpha X/2}}+\,O\!\left(e^{-\lambda X}\right).
\label{n1TFfin}
\end{equation}
Estimating the argument of the arctangent here, we have
\begin{equation}
\sqrt{\frac{\alpha}{2X}}=\sqrt{\frac{\hbar^2\tilde{k}_z^2}{4M_{ex}2\gamma_0X}}=\frac{\pi w}{3}\sqrt{\frac{\hbar^2}{M_{ex}b^2}}\frac{1}{\sqrt{2\gamma_0X}}\gtrsim4
\label{estimate1}
\end{equation}
for all reasonable $X\!\lesssim0.1\,\mbox{eV}/2\gamma_0$. We use $\tilde{k}_z\!=\!2\pi w/3b$ with $1\!\le\!w\!\le\!\sqrt{3}$ to cover~all possible CN chiralities, and $M_{ex}$ is taken to be twice the free electron mass.~Thus, Eq.~(\ref{n1TFfin}) can be further approximated using the large argument expansion for the arctangent ($\sim\!\pi/2$), to result in $n_1\!\approx\!\pi/(2\lambda\sqrt{\alpha X/2})$ (exponential smallness neglected), yielding the maximum mean population for the upper-branch exciton plasmons as follows
\begin{equation}
n_1(T_c)\approx\frac{3}{2w}\sqrt{\frac{M_{ex}b^2}{\hbar^2}}\sqrt{k_BT_c(F)}\,.
\label{n1Tc}
\end{equation}

Next, using Eq.~(\ref{nk}), we proceed to calculate the mean population $\Delta n_1$ of the upper-branch exciton-plasmons with $k_z\!\ne\!0$. We have
\[
\Delta n_1=\frac{2}{N}\sum_{s=1}^m\left[\sum_{k_z=-\tilde{k}_z}^{-2\pi/L}\!\!\!\langle n_1(s,k_z)\rangle+
\!\!\!\!\!\sum_{k_z=2\pi/L}^{\tilde{k}_z}\!\!\!\!\langle n_1(s,k_z)\rangle\right]
\]\vspace{-0.5cm}
\[
=\frac{2}{N}\sum_{s=1}^m\frac{L}{\pi}\!\int_{2\pi/L}^{\tilde{k}_z}\!\!\!\!\!dk_z\,\langle n_1(s,k_z)\rangle=\frac{1}{m}\sum_{s=1}^m\int_{t_0}^{1}\!\frac{dt}{e^{\lambda x_1(s,t)}-1}\,,
\]
where $t_0\!=\!2\pi/L\tilde{k}_z\!=\!4m/N$. After using the (geometric) series expansion followed by term-by-term integration, we arrive at
\begin{equation}
\Delta n_1=\frac{\sqrt{\pi}}{2}\sum_{n=0}^{+\infty}\frac{e^{-(n+1)\lambda X}}{\sqrt{(n+1)\lambda\alpha/2}}
\label{Dn1TF}
\end{equation}\vspace{-0.5cm}
\[
\times\left\{\mbox{Erf}\!\left[\sqrt{(n+1)\lambda\alpha/2}\,\right]-\mbox{Erf}\!\left[t_0\sqrt{(n+1)\lambda\alpha/2}\,\right]\right\}.
\]
This, after using Eq.~(\ref{summationformula}) to sum up the series, results in
\begin{equation}
\Delta n_1\approx\frac{1}{\lambda\sqrt{\alpha X/2}}\left[\,\arctan\!\left(\!\sqrt{\alpha/2X}\right)\right.
\label{Dn1TFfin}
\end{equation}\vspace{-0.5cm}
\[
\left.-\arctan\!\left(\!t_0\sqrt{\alpha/2X}\right)\right]+\,O\!\left(e^{-\lambda X}\right).
\]
Here, the estimate for the first term argument is given by Eq.~(\ref{estimate1}). The second term argument is then $\sim\!t_0\!\sim\!1/N$, so that $\arctan(t_0\sqrt{\alpha/2X})\!\sim\!0$ is negligible.~Neglecting also exponentially small terms, as we do in Eq.~(\ref{n1Tc}), brings Eq.~(\ref{Dn1TFfin}) to the following form
\begin{equation}
\Delta n_1(T,F)\approx\frac{3}{2w}\sqrt{\frac{M_{ex}b^2}{\hbar^2}}\frac{k_BT}{\sqrt{k_BT_c(F)}}\,.
\label{Dn1TFfinfin}
\end{equation}

Finally, using Eqs.~(\ref{n1Tc}) and (\ref{Dn1TFfinfin}), we obtain the (BEC) fraction of zero-momentum exciton-plasmons at $T\!\le\!T_c(F)$, defined as
\[
\langle n_{10}\rangle(T\!\le\!T_c,F)=\frac{n_1(T_c)-\Delta n_1(T,F)}{n_1(T_c)}\,,
\]
in the form as given by Eq.~(\ref{n10}).

\section{~Derivation of Eq.~(\ref{n10ep})}\label{appB}

General theory of the exciton-plasmon interactions in individual CNs (see Ref.\cite{Bondarev09}) relates the operators $\hat{\xi}^\dag_\mu(\mathbf{k})$ and $\hat{\xi}_\mu(\mathbf{k})$ that create and annihilate, respectively, exciton-plasmons of branch $\mu\;(=\!1,2)$ with the momentum $\mathbf{k}$, to the exciton creation-annihilation operators $B_{\mathbf{k},f}^\dag$, $B_{\mathbf{k},f}$ [$f$-internal state with the energy $E_f(\mathbf{k})$ used in Eq.~(\ref{x12})] and the plasmon creation-annihilation operators $\hat{f}^\dag(\mathbf{k},\omega)$, $\hat{f}(\mathbf{k},\omega)$ as follows
\begin{eqnarray}
\hat{\xi}_\mu^\dag(\mathbf{k})=\sum_{f}\left[u_{\mu f}^{(ex)}B_{\mathbf{k},f}^\dag-v_{\mu f}^{\ast\,(ex)}B_{-\mathbf{k},f}\right]\hskip0.5cm\label{xik}\\
+\int_0^\infty\!\!\!\!\!d\omega\left[u_\mu^{\ast\,(p)}(\omega)\hat{f}^\dag(\mathbf{k},\omega)-v_\mu^{(p)}(\omega)\hat{f}(-\mathbf{k},\omega)\right],\nonumber
\end{eqnarray}\vspace{-0.5cm}
\[
\hat{\xi}_\mu(\mathbf{k})=[\hat{\xi}^\dag_\mu(\mathbf{k})]^\dag.
\]
Here, $u_{\mu f}^{(ex)}\!$, $v_{\mu f}^{\ast\,(ex)}\!$, $u_\mu^{\ast\,(p)}\!$ and $v_\mu^{(p)}\!$ are the complex mixing coefficients that define the (Bogoliubov) unitary canonical transformation on the total Hamiltonian of the coupled exciton-plasmon system to bring it to the diagonal form [used in Eq.~(\ref{n})].~These mixing coefficients are given by the solutions to the following set of simultaneous linear equations
\begin{eqnarray}
\left(\hbar\omega_\mu\!-\!E_f\right)u^{(ex)}_{\mu f}\!\!=\!i\!\!\int_0^\infty\!\!\!\!\!\!\!d\omega D_{\!f}(\omega)\!\left[u_\mu^{\ast\,(p)}(\omega)\!-\!v^{(p)}_\mu(\omega)\right]\!,\hskip0.5cm\label{eq1}\\
\left(\hbar\omega_\mu\!+\!E_f\right)v_{\mu f}^{\ast\,(ex)}\!\!=-i\!\!\int_0^\infty\!\!\!\!\!\!\!d\omega D_{\!f}(\omega)\!\left[u_\mu^{\ast\,(p)}(\omega)\!-\!v^{(p)}_\mu(\omega)\right]\!,\hskip0.2cm\label{eq2}\\
\hbar\!\left(\omega_\mu\!-\omega\right)u_\mu^{\ast\,(p)}(\omega)=-i\!\sum_f\!D_{\!f}(\omega)\!\left[u^{(ex)}_{\mu f}\!+\!v_{\mu f}^{\ast\,(ex)}\right]\!,\hskip0.5cm\label{eq3}\\
\hbar\!\left(\omega_\mu\!+\omega\right)v^{(p)}_\mu(\omega)=-i\!\sum_f\!D_{\!f}(\omega)\!\left[u^{(ex)}_{\mu f}\!+\!v_{\mu f}^{\ast\,(ex)}\right]\!,\hskip0.7cm\label{eq4}
\end{eqnarray}
where $D_{\!f}(\omega)\!=\!\hbar\sqrt{\Gamma_0^f(\omega)\rho(\omega)/2\pi}$ is the exciton-plasmon interaction matrix element with $\Gamma_0^f(\omega)$ and $\rho(\omega)$ representing the frequency dependences of the exciton spontaneous decay rate and that of the plasmon DOS function responsible for the exciton decay rate variation due to its (non-radiative) coupling to nanotube plasmon modes.

Equations~(\ref{eq1})-(\ref{eq4}) define the mixing coefficients, as well as they define the dispersion relation for the exciton-plasmon energy $\hbar\omega_\mu(\textbf{k})$ that is used in Eqs.~(\ref{n})-(\ref{nk}) and (in the dimensionless form) in Eqs.~(\ref{x12}) and (\ref{x1fst}). The explicit form of the dispersion relation can be found in Ref.\cite{Bondarev09}.~Here, we derive the solutions for the mixing coefficients.

We are particularly interested in finding the coefficients $u_{\mu f}^{(ex)}$ and $u_\mu^{\ast\,(p)}(\omega)$ since, as we can see from Eq.~(\ref{xik}), the absolute value squared of their ratio shows the exciton participation against the plasmon participation in an exciton-plasmon excitation created.~Combining Eqs.~(\ref{eq1}) and (\ref{eq2}) yields
\begin{equation}
v_{\mu f}^{\ast\,(ex)}=\frac{E_f-\hbar\omega_\mu}{E_f+\hbar\omega_\mu}\,u^{(ex)}_{\mu f},
\label{eq5}
\end{equation}
while from Eqs.~(\ref{eq3}) and (\ref{eq4}) we have
\begin{equation}
v_\mu^{(p)}(\omega)=\frac{\omega_\mu-\omega}{\omega_\mu+\omega}\,u^{\ast\,(p)}_\mu(\omega)\,.
\label{eq6}
\end{equation}
Next, from Eq.~(\ref{eq3}), using Eq.~(\ref{eq5}) in its right-hand side, one obtains
\begin{equation}
u_\mu^{\ast\,(p)}(\omega)=i\sum_f\frac{2E_fD_{\!f}(\omega)}{\hbar(\omega-\omega_\mu)(E_f+\hbar\omega_\mu)}\,u^{(ex)}_{\mu f},
\label{eq7}
\end{equation}
which being substituted into Eq.~(\ref{eq6}) results in
\begin{equation}
v_\mu^{(p)}(\omega)=-i\sum_f\frac{2E_fD_{\!f}(\omega)}{\hbar(\omega+\omega_\mu)(E_f+\hbar\omega_\mu)}\,u^{(ex)}_{\mu f}.
\label{eq8}
\end{equation}

Equations (\ref{eq5}), (\ref{eq7}) and (\ref{eq8}), with $u^{(ex)}_{\mu f}$ determined by the normalization condition, solve the equations set (\ref{eq1})-(\ref{eq4}).~They can be written in the dimensionless variables used throughout this paper as follows
\begin{eqnarray}
v_{\mu f}^{\ast\,(ex)}=\frac{\varepsilon_f-x_\mu}{\varepsilon_f+x_\mu}\,u^{(ex)}_{\mu f},\hskip2.1cm\label{soldl1}\\
\bar{u}_\mu^{\ast\,(p)}(x)=i\sum_f\frac{2\,\varepsilon_f\sqrt{\bar\Gamma_0^f(x)\rho(x)/2\pi}}{(x-x_\mu)(\varepsilon_f+x_\mu)}\,u^{(ex)}_{\mu f},\hskip0.7cm\label{soldl2}\\
\bar{v}_\mu^{(p)}(x)=-i\sum_f\frac{2\,\varepsilon_f\sqrt{\bar\Gamma_0^f(x)\rho(x)/2\pi}}{(x+x_\mu)(\varepsilon_f+x_\mu)}\,u^{(ex)}_{\mu f},\hskip0.7cm\label{soldl3}
\end{eqnarray}
where $\bar{u}_\mu^{\ast\,(p)}\!\!=\!u_\mu^{\ast\,(p)}\!\sqrt{2\gamma_0/\hbar}$ and $\bar{v}_\mu^{(p)}\!\!=\!v_\mu^{(p)}\!\sqrt{2\gamma_0/\hbar}$ are the dimensionless counterparts of the corresponding mixing coefficients.~Assuming further that the ground internal state of the exciton contributes the most to the summations over $f$ in the expressions above, we arrive at the ratio of interest in the form ($f$-subscript dropped)
\begin{equation}
\frac{\left|u^{(ex)}_\mu\right|^2}{\left|\bar{u}_\mu^{\ast\,(p)}\right|^2}\approx\frac{\pi(x-x_\mu)^2(1+x_\mu/\varepsilon)^2}{2\bar\Gamma_0(x)\rho(x)}\,.
\label{ratio}
\end{equation}

To obtain the exciton participation rate in the upper branch ($\mu\!=\!1$) exciton-plasmon BEC population fraction, we note that the denominator in Eq.~(\ref{ratio}) is nothing but the (dimensionless) exciton-plasmon interaction matrix element squared. This is only non-zero when the exciton energy $\varepsilon\!=\!\varepsilon(F,t)$ and the plasmon resonance energy $x_p(F)$ are close in their values.~As this takes place, the plasmon DOS $\rho(x)$ can be legitimately approximated by the Lorentzian of the half-width-at-half-maximum $\Delta x_p$ (representing the inverse plasmon life-time) of the form
\[
\rho(x)\approx\frac{\rho(x_p)\Delta x_p^2}{(x-x_p)^2+\Delta x_p^2}\,,
\]
in which the frequency $x$ is equal to $\varepsilon(F,t)$ and this latter one is assumed to be of the order of $x_p(F)$.~With this in mind, we write the denominator in Eq.~(\ref{ratio}) as follows
\[
2\bar\Gamma_0(x)\rho(x)=2\bar\Gamma_0[\varepsilon(F,t)]\rho[\varepsilon(F,t)]=\frac{X^2(F,t)}{\Delta x_p}\,.
\]
Then, Eq.~(\ref{ratio}) for the upper branch exciton-plasmons takes the form
\begin{equation}
\frac{\left|u^{(ex)}_1\right|^2}{\left|\bar{u}_1^{\ast\,(p)}\right|^2}\approx\frac{\pi\Delta x_p(x-x_1)^2[1+x_1/\varepsilon(F,t)]^2}{X^2(F,t)}\,,
\label{ratio1}
\end{equation}
where the energies should be counted from $x_2(F,t\!=\!0)$ as per our previous convention.

Figure~\ref{fig3}~(a) shows functions $2\gamma_0X(F,t)/k_B$ $[=\!T_c(F,t)]$ we calculated for the lowest bright ground-internal-state excitons coupled to the nearest inter-band plasmons in the four CNs of our choice here.~Function $X(F,t)$ is sharp resonance shaped with the peak position determined by the condition $\varepsilon(F,t)\!=\!x_p(F)$ [cf. Fig.~\ref{fig2}~(b)]. As $F$ increases, the peak shifts down to $t\!\sim\!0$, yielding $X(F)\!=\!X(F,t\!=\!0)$ sharply peaked around $F\!=\!F_c$. This suggests that, when at resonance, the ratio (\ref{ratio1}) is generally much less than one, so that exciton-plasmon excitations are dominated by plasmons. However, a slight detuning from the resonance condition decreases $X^2(F,t)$ dramatically, making the ratio (\ref{ratio1}) dramatically increase and excitons dominate an exciton-plasmon state.

Using Eq.~(\ref{ratio1}), the exciton participation rate in the exciton-plasmon BEC population fraction can be found as follows
\[
\langle n_{10}\rangle_{\!_{E\!/\!P}}(x,T,F)=\left.\frac{\left|u^{(ex)}_1\right|^2}{\left|\bar{u}_\mu^{\ast\,(p)}\right|^2}\,\right|_{t=0}\!\!\!\langle n_{10}\rangle(T\!\le\!T_c,F)\,,
\]
which brings us to Eq.~(\ref{n10ep}).


\begin{thebibliography}{99}
\bibitem{endnote}Electrostatic fields reported for the QCSE on CNs must be multiplied by $1/\sqrt{4\pi\epsilon_0}$ to result in field strengths $\sim$V/nm, consistent with fields observed for the QCSE on bilayer graphene~\cite{Fengwang09}. I.V.B. is indebted to Dr. A.Hoegele of LMU, Germany, for bringing this issue to his attention.
\end{thebibliography}
\end{document}